\documentclass[pre,showpacs]{revtex4}
\usepackage{graphicx}

\begin{document}

\title{Two-point correlation function of the fractional Ornstein-Uhlenbeck process}

\author{A. Baule$^1$ and R. Friedrich$^2$}

\affiliation{                    
$^1$School of Physics and Astronomy, University of Leeds, LS2 9JT, United Kingdom\\
$^2$Institute for Theoretical Physics, University of M\"unster, Wilhelm-Klemm Str. 9, 48149 M\"unster, Germany
}

\begin{abstract}

We calculate the two-point correlation function $\langle x(t_2)x(t_1)\rangle$ for a subdiffusive continuous time random walk in a parabolic potential, generalizing well-known results for the single-time statistics to two times. A closed analytical expression is found for initial equilibrium, revealing non-stationarity and a clear deviation from a Mittag-Leffler decay. Our result thus provides a new criteria to assess whether a given stochastic process can be identified as a continuous time random walk.

\end{abstract}


\maketitle

\section{Introduction}

During the last decades, continuous time random walks (CTRWs) have been used to model a wide variety of stochastic processes which can not be described by the usual type of Langevin
equations. Especially, CTRW statistics have been successfully applied to systems exhibiting non-Gaussian probability distributions. For a review we refer the reader to the comprehensive review articles by Metzler and Klafter \cite{Metzler1,Metzler2}. It has become evident however, that due to their non-Markovian character, assessing the probability distributions at a single time is insufficient in order to determine the nature of the underlying stochastic process. To this end, multiple time probability distributions of CTRWs have been investigated in a recent line of research \cite{Allegrini,Barsegov,Sanda,Baule1,Baule2,Barkai1}. These probability distributions can be used to determine multiple time correlation functions, which in turn allow for a comparison with experimental data. Experiments in which such correlation functions play a crucial role have recently been reported in the context of protein conformational dynamics \cite{Yang,Min}, and blinking nanocrystals \cite{Margolin}.

The purpose of the present paper is to provide the two-time correlation function for a subdiffusive CTRW-process of a particle in a harmonic potential, also referred to as fractional Ornstein-Uhlenbeck process. A general form of this correlation function as well as a closed analytical expression for initial equilibrium position of the particle are derived as our main results. It is clear from many textbook examples in physics, that motion in a parabolic potential is of utmost importance, and has been used as an approximation e.g. in \cite{Yang}. Our results could therefore be readily compared with data from existing and future experiments.

The remainder of this paper is organized as follows. In the next section we present the basic equations for the CTRW model. Then, well-known results for the single-time statistics are rederived. The two main sections follow, in which we derive our two main results, the general expression for the correlation function and a special analytic form. We conclude with a few remarks.

\section{\label{sec_basic}Coupled Langevin equations of the fractional Ornstein-Uhlenbeck process}

In this section we briefly review a well-known representation of CTRWs in terms of coupled Langevin equations \cite{Fogedby}. In the continuum limit of infinitesimal step lengths, the one dimensional motion of a random walk particle in a parabolic potential $V(x)=1/2\,\gamma \,x^2$ is described as:
\begin{eqnarray} \frac{dX(s)}{ds}&=& -\gamma X(s) + \eta(s),
\label{Langevin1}\\ \frac{dt(s)}{ds}&=& \tau(s). \label{Langevin2}
\end{eqnarray}
In this framework the CTRW is parametrized by the continuous path 
variable $s$, which may be regarded as arclength along the trajectory. $X$ denotes the physical space and $t$ the physical time. Both are given as stochastic processes in the 'eigentime' $s$.
Their statistics are determined by the properties of the stochastic variables $\eta(s)$ and $\tau(s)$. We only consider the special case of statistically independent increments $\eta$ and $\tau$, corresponding to uncoupled jump lengths and waiting times. Assuming $\eta(s)$ as a standard Langevin force with properties $\left<\eta(s)\right>=0$ and $\left<\eta(s)\eta(s')\right>=2\sigma\delta(s-s')$, we see that $X(s)$ accordingly corresponds to the ordinary Markovian Ornstein-Uhlenbeck process \cite{Risken}. CTRW characteristics enter via the process $t(s)$. Its increments $\tau(s)$ are assumed to be broadly distributed such that $t(s)$ represents an asymmetric L\'evy-stable process of order $\alpha$ with $0<\alpha<1$. L\'evy-stable processes of this kind induce a diverging characteristic waiting time $\left<t(s)\right>$. As a result, the second moment of the corresponding free diffusion process, i.e. setting the force term in Eq.~(\ref{Langevin1}) to zero, yields subdiffusive scaling $\langle x^2(t)\rangle\sim t^\alpha$. Therefore we refer to the stochastic process specified by Eqs.~(\ref{Langevin1})---(\ref{Langevin2}) as a subdiffusive CTRW in a parabolic potential, or fractional Ornstein-Uhlenbeck process \cite{Metzler1}.

The CTRW in physical time $t$ is determined by the subordinated process $X(t)=X(s(t))$, where $s(t)$ is an inverse L\'evy-stable process \cite{Bingham}. We define the probability distributions (pdfs) of $X(t)$ as
\begin{eqnarray}
\label{2_pdf_def}
f(x_2,t_2;x_1,t_1)&=&\langle\delta(x_2-X(s_2))\delta(s_2-s(t_2))\delta(x_1-X(s_1))\delta(s_1-s(t_1))\rangle,
\end{eqnarray}
and accordingly for multiple times. It is a basic result, that these $n$-point pdfs can be expressed as an integral transformation \cite{Baule1}
\begin{eqnarray}
\label{path_integral}
f(\{x_i,t_i\})&=&\int_0^\infty ds_1...\int_0^\infty ds_n\;h(\{s_i,t_i\})f_M(\{x_i,s_i\}),\nonumber\\
\end{eqnarray}
here $h(\{s_i,t_i\})$ denotes the $n$-point pdf of the process $s(t)$ and $f_M(\{x_i,s_i\})$ the $n$-point pdf of the Markovian process $X(s)$. Equation~(\ref{path_integral}) states that the pdfs of the fractional Ornstein-Uhlenbeck process are determined by a transformation of the corresponding Markovian pdfs. The integral kernel is generated by an inverse L\'evy-stable process. From this integral transformation it is obvious, that all moments of the process $X(t)$ are obtained by averaging the moments of the Markovian process $X(s)$ with the subordinator $s(t)$. In the single and two point case, $h$ assumes a simple analytical form in Laplace space.

\section{\label{sec_single}Single-time moments}

Let us first consider the well-known single-time moments $\langle x(t) \rangle$ and $\langle x^2(t) \rangle$. In this case Eq.~(\ref{path_integral}) leads to
\begin{eqnarray}
\langle x^n(t) \rangle=\int_0^\infty ds\,\langle x^n(s) \rangle\, h(s,t).
\end{eqnarray}
Here the single-time pdf $h(s,t)$ is given as \cite{Barkai2}
\begin{eqnarray}
h(s,t)=\frac{1}{\alpha}\frac{t}{s^{1+1/\alpha}}L_\alpha\left(\frac{t}{s^{1/\alpha}}\right).
\end{eqnarray}
$L_\alpha$ denotes a single-sided L\'evy distribution of order $\alpha$. As mentioned above, calculations involving $h$ are more conveniently performed in Laplace space. We define Laplace transforms in the usual way: $\mathcal{L}\{g(t)\}=\int_0^\infty g(t)e^{-\lambda t}dt$, accordingly for multiple variables $t_i\rightarrow\lambda_i$. Functions with Laplace space argument always denote the Laplace transform if not otherwise indicated: $g(\lambda)\equiv\mathcal{L}\{g(t)\}$. In Laplace space $h(s,t)$ reads $h(s,\lambda)=\lambda^{\alpha-1}e^{-\lambda^\alpha s}$ and with $\langle x(s)\rangle=x_0e^{-\gamma s}$ (see e.g. Ch. 3 in \cite{Risken}) we can write the Laplace transform of $\langle x(t)\rangle$
\begin{eqnarray}
\langle x(\lambda)\rangle&=&x_0\int_0^\infty ds\,\lambda^{\alpha-1}e^{-(\gamma+\lambda^\alpha)s}=x_0\frac{\lambda^{\alpha-1}}{\gamma+\lambda^\alpha}.
\end{eqnarray}
Performing a series expansion of $1/(\gamma+\lambda^\alpha)$ and term by term Laplace-inversion, the final result can be expressed with the help of the one-parameter Mittag-Leffler function $E_\alpha$ \cite{Podlubny}
\begin{eqnarray}
\mathcal{L}^{-1}\left\{\frac{\lambda^{\alpha-1}}{\gamma+\lambda^\alpha}\right\}=\sum_{n=0}^\infty\frac{(-\gamma t^\alpha)^n}{\Gamma(n\alpha+1)}\equiv E_\alpha(-\gamma t^\alpha),
\end{eqnarray}
and therefore
\begin{eqnarray}
\label{st_firstmoment}
\langle x(t)\rangle=x_0 E_\alpha(-\gamma t^\alpha).
\end{eqnarray}
Likewise we can calculate the second moment $\langle x^2(t)\rangle$. Note that $\langle x^2(s)\rangle=(x_0^2-\sigma/\gamma)$ $\exp(-2\gamma s)+\sigma/\gamma$ (\cite{Risken}, Ch. 3), and we obtain the result
\begin{eqnarray}
\label{moment2}
\langle x^2(t)\rangle=\left(x_0^2-\frac{\sigma}{\gamma}\right)E_\alpha(-2\gamma t^\alpha)+\frac{\sigma}{\gamma}.
\end{eqnarray}
In the long-time limit the mean-squared displacement approaches the thermal equilibrium value
$\langle x^2\rangle_{eq}=\sigma/\gamma=k_BT/\gamma$. Above results for the single-time moments have been derived in the context of the fractional Fokker-Planck equation \cite{Metzler3}.

\section{\label{sec_corr}Correlation function}

The calculation of two-time moments is more involved but follows the same procedure. In the following we restrict our consideration to the simplest case $\langle x(t_2)x(t_1)\rangle$, higher order moments can be obtained along the same lines. With the help of the integral transformation Eq.~(\ref{path_integral}), we can express the correlation function as before
\begin{eqnarray}
\label{corr_def}
\langle x(t_2)x(t_1)\rangle&=&\int_0^\infty ds_2\int_0^\infty ds_1\,\langle x(s_2)x(s_1)\rangle h(s_2,t_2;s_1,t_1).
\end{eqnarray} 
Here, the correlation function $\langle x(s_2)x(s_1)\rangle$ of the standard Ornstein-Uhlenbeck process is well known \cite{Risken} and follows from the Langevin equation (\ref{Langevin1}) in a straightforward way
\begin{eqnarray}
\label{OU_corr}
\langle x(s_2)x(s_1)\rangle&=&\left(x_0^2-\frac{\sigma}{\gamma}\right)e^{-\gamma(s_2+s_1)}+\frac{\sigma}{\gamma} e^{-\gamma|s_2-s_1|)}.\nonumber\\
\end{eqnarray}
On the other hand, the two-time pdf $h(s_2,t_2;s_1,t_1)$ has a closed analytical form in Laplace space only \cite{Baule1}
\begin{eqnarray}
\label{pdf_h2}
h(s_2,\lambda_2;s_1,\lambda_1)&=&\frac{1}{\lambda_2\lambda_1}\frac{\partial^2}{\partial s_2\partial s_1}\left[\Theta(s_2-s_1)e^{-(\lambda_1+\lambda_2)^\alpha s_1-\lambda_2^\alpha (s_2-s_1)}\right.\nonumber\\
&&+\Theta(s_1-s_2)\left.e^{-(\lambda_1+\lambda_2)^\alpha s_2-\lambda_1^\alpha (s_1-s_2)}\right].
\end{eqnarray}
Performing the integrations in Eq.~(\ref{corr_def}) with these two expressions, we obtain the following result for the correlation function in Laplace space
\begin{eqnarray}
\langle x(\lambda_2)x(\lambda_1)\rangle&=&\sigma(\lambda_1+\lambda_2)^{-\alpha}\left[\frac{1}{\lambda_1}\frac{\lambda_2^{\alpha-1}}{(\gamma+\lambda_2^\alpha)}+\frac{1}{\lambda_2}\frac{\lambda_1^{\alpha-1}}{(\gamma+\lambda_1^\alpha)}\right]\nonumber\\
&&+\left(x_0^2-\frac{\sigma}{\gamma}\right)\frac{\gamma^2}{\lambda_1 \lambda_2(2\gamma+(\lambda_1+\lambda_2)^\alpha)}\left[\frac{1}{\gamma+\lambda_2^\alpha}+\frac{1}{\gamma+\lambda_1^\alpha}\right]\nonumber\\
&&+x_0^2\left[\frac{1}{\lambda_2}\frac{\lambda_1^{\alpha-1}}{(\gamma+\lambda_1^\alpha)}+\frac{1}{\lambda_1}\frac{\lambda_2^{\alpha-1}}{(\gamma+\lambda_2^\alpha)}-\frac{1}{\lambda_1\lambda_2}\right].
\end{eqnarray}
Laplace inversion yields our first main result, the correlation function in physical time
\begin{eqnarray}
\label{corr_result}
\langle x(t_2)x(t_1)\rangle
&=&\sigma\left(\frac{\partial}{\partial t_2}+\frac{\partial}{\partial t_1}\right)^{-\alpha}[E_\alpha(-\gamma t_2^\alpha)+E_\alpha(-\gamma t_1^\alpha)]+\left(x_0^2-\frac{\sigma}{\gamma}\right)Z_s(\gamma,t_2;\gamma,t_1)\nonumber\\
&&+\frac{\sigma}{\gamma}[E_\alpha(-\gamma t_2^\alpha)+E_\alpha(-\gamma t_1^\alpha)-1].
\end{eqnarray}
Here we have introduced the two-time fractional integral operator $(\partial_{t_2}+\partial_{t_1})^{-\alpha}$, which can be represented as \cite{Baule1}
\begin{eqnarray}
\left(\frac{\partial}{\partial t_2}+\frac{\partial}{\partial t_1}\right)^{-\alpha}g(t_2,t_1)=\frac{1}{\Gamma(\alpha)}\int_0^{Min(t_2,t_1)}dt\,t^{\alpha-1}
g(t_2-t,t_1-t),
\end{eqnarray}
and whose Laplace transform reads
\begin{eqnarray}
\mathcal{L}\left\{\left(\frac{\partial}{\partial t_2}+\frac{\partial}{\partial t_1}\right)^{-\alpha}g(t_2,t_1)\right\}=(\lambda_2+\lambda_1)^{-\alpha}g(\lambda_2,\lambda_1).\nonumber\\
\end{eqnarray}
Furthermore, Eq.~(\ref{corr_result}) contains the Laplace transform of the two-time pdf $h$ of the process $s(t)$
\begin{eqnarray}
Z_s(u_2,t_2;u_1,t_1)&\equiv&\langle e^{-u_2s(t_2)-u_1s(t_1)}\rangle\nonumber\\
&=&\int_0^\infty ds_2 \int_0^\infty ds_1\,e^{-u_2s_2-u_1s_1}h(s_2,t_2;s_1,t_1), 
\end{eqnarray}
evaluated at $u_2=u_1=\gamma$. The occurence of this Laplace transform is simply a consequence of the first term in Eq.~(\ref{OU_corr}) and the transformation Eq.~(\ref{corr_def}). In the following we use the properties of the pdf $h(s_2,t_2;s_1,t_1)$ in order to validate certain limit cases of Eq.~(\ref{corr_result}).

\subsection{The special case $t_1=0$}
We first rederive the autocorrelation function $\langle x(t)x(0)\rangle$ from Eq.~(\ref{corr_result}). For $t_1=0$ we know that $h(s_2,t_2;s_1,0)=h(s_2,t_2)\delta(s_1)$ and thus $Z_s(\gamma,t_2;\gamma,0)=Z_s(\gamma,t_2)=E_\alpha(-\gamma t_2^\alpha)$. The fractional integral operator has zero contribution for $t_1=0$. We therefore recover the well known result (see Eq.~(\ref{st_firstmoment}))
\begin{eqnarray}
\label{corr_t0}
\langle x(t)x(0)\rangle=x_0^2E_\alpha(-\gamma t^\alpha)=x_0\langle x(t)\rangle.
\end{eqnarray}

\subsection{The limit $t_2\rightarrow t_1$}
For $t_2\rightarrow t_1$ Eq.~(\ref{corr_result}) should reduce to the quadratic moment Eq.~(\ref{moment2}). In this limit the two-time fractional integral reduces to the familiar Riemann-Liouville fractional integral
\begin{eqnarray}
\lim_{t_2\rightarrow t_1}\left(\frac{\partial}{\partial t_2}+\frac{\partial}{\partial t_1}\right)^{-\alpha}E_\alpha(-\gamma t_1^\alpha)&=&{_0}D_{t_1}^{-\alpha}E_\alpha(-\gamma t_1^\alpha)\nonumber\\
&=&\frac{1}{\Gamma(\alpha)}\int_0^{t_1}dt\,t^{\alpha-1}E_\alpha(-\gamma (t_1-t)^\alpha).
\end{eqnarray}
Using the series representation of the Mittag-Leffler function, the convolution integral of each term is calculated in Laplace space in a straightforward way
\begin{eqnarray}
\mathcal{L}\{{_0}D_{t_1}^{-\alpha}E_\alpha(-\gamma t_1^\alpha)\}
&=&-\frac{1}{\gamma}\sum_{n=1}^\infty\frac{1}{\lambda_1}(-\gamma\lambda_1^{-\alpha})^n.
\end{eqnarray}
Therefore
\begin{eqnarray}
\label{fracInt_ML}
{_0}D_{t_1}^{-\alpha}E_\alpha(-\gamma t_1^\alpha)=\frac{1}{\gamma}\left(1-E_\alpha(-\gamma t_1^\alpha)\right).
\end{eqnarray}
Further note that $\lim_{t_2\rightarrow t_1}h(s_2,t_1;s_1,t_1)=\delta(s_2-s_1)h(s_1,t_1)$. A quick calculation then yields $Z_s(\gamma,t_1;\gamma,t_1)=E_\alpha(-2\gamma t_1^\alpha)$. Substituting this expression and Eq.~(\ref{fracInt_ML}) into Eq.~(\ref{corr_result}) leads to the result $\lim_{t_2\rightarrow t_1}\langle x(t_2)x(t_1)\rangle=\langle x^2(t_1) \rangle$.

\subsection{The limit $\alpha\rightarrow 1$}
In the limit $\alpha\rightarrow 1$ the first moment of the CTRW waiting time distribution is finite and we expect that Eq.~(\ref{corr_result}) yields the standard correlation Eq.(\ref{OU_corr}). This can be seen as follows. The pdf $h$ factorizes into delta-functions in the $\alpha\rightarrow 1$ limit: $\lim_{\alpha\rightarrow 1}h(s_2,t_2;s_1,t_1)=\delta(s_2-t_2)\delta(s_1-t_1)$ (see Eq.~(\ref{pdf_h2})), and thus $\lim_{\alpha\rightarrow 1}Z_s(\gamma,t_2;\gamma,t_1)=e^{-\gamma(t_2+t_1)}$. In the same limit the Mittag-Leffler function reduces to the usual exponential function and the two-time fractional integral to a normal integral. The result of this calculation is
\begin{eqnarray}
\left(\frac{\partial}{\partial t_2}+\frac{\partial}{\partial t_1}\right)^{-1}\left(e^{-\gamma t_2}+e^{-\gamma t_1}\right)&=&\Theta(t_2-t_1)\frac{1}{\gamma}\left(1+e^{-\gamma(t_2-t_1)}-e^{-\gamma t_2}-e^{-\gamma t_1}\right)\nonumber\\
&&+\Theta(t_1-t_2)\frac{1}{\gamma}\left(1+e^{-\gamma(t_1-t_2)}-e^{-\gamma t_2}-e^{-\gamma t_1}\right).\nonumber\\
\end{eqnarray}
Substituting these expressions into Eq.~(\ref{corr_result}) yields the standard correlation of the Ornstein-Uhlenbeck process Eq.~(\ref{OU_corr}) as expected.

\section{\label{sec_init}Initial equilibrium position}

\begin{figure}
\begin{center}
\includegraphics[width=8cm]{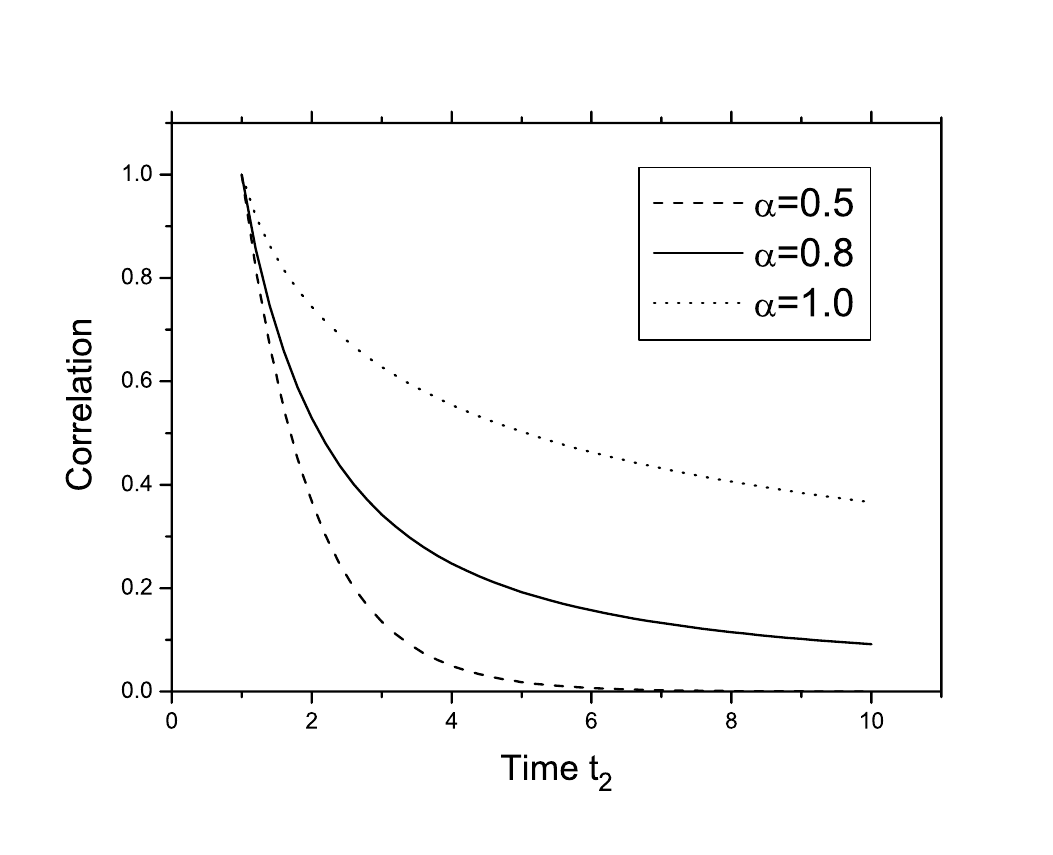}
\caption{\label{Fig_corr}The correlation function Eq.~(\ref{corr_eq2}) for $\gamma=\sigma=1$, $t_2>t_1=1$ and three different $\alpha$-values: $\alpha=1.0$ (dashed line), $\alpha=0.8$ (solid), and $\alpha=0.5$ (dotted). For $\alpha=1$ the decay is exponential, see Eq.~(\ref{corr_a1}).}
\end{center}
\end{figure}

\begin{figure}
\begin{center}
\includegraphics[width=8cm]{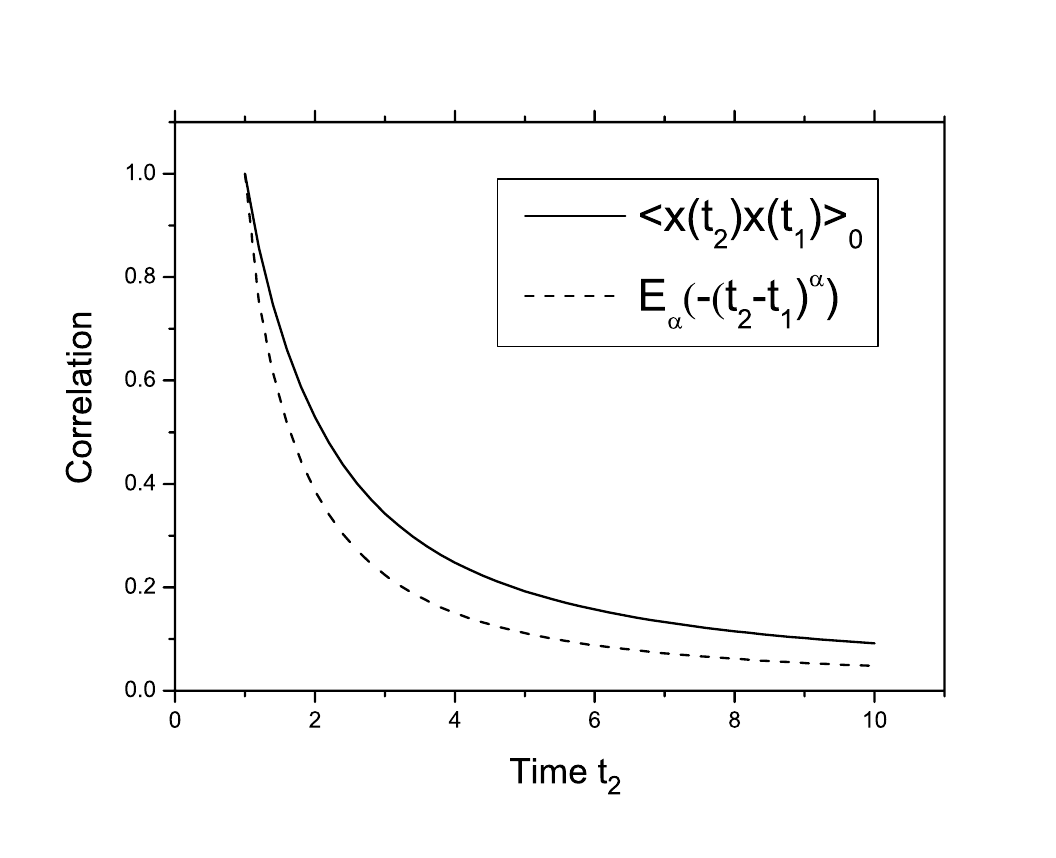}
\caption{\label{Fig_corrML}The correlation function Eq.~(\ref{corr_eq2}) for $\gamma=\sigma=1$, $t_2>t_1=1$ and $\alpha=0.8$ compared with the Mittag-Leffler function $E_{\alpha}(-(t_2-t_1)^{\alpha})$ for same parameter values.}
\end{center}
\end{figure}

If we choose as special initial value the thermal equilibrium displacement $\langle x^2 \rangle_{eq}$, we can determine a closed analytical expression for the correlation function $\langle x(t_2)x(t_1) \rangle$. For $x_0^2=\sigma/\gamma$ Eq.~(\ref{corr_result}) leads to
\begin{eqnarray}
\label{corr_eq}
\langle x(t_2)x(t_1)\rangle_0&=&\sigma\left(\frac{\partial}{\partial t_2}+\frac{\partial}{\partial t_1}\right)^{-\alpha}[E_\alpha(-\gamma t_2^\alpha)+E_\alpha(-\gamma t_1^\alpha)]\nonumber\\
&&+\frac{\sigma}{\gamma}[E_\alpha(-\gamma t_2^\alpha)+E_\alpha(-\gamma t_1^\alpha)-1].
\end{eqnarray} 
Here, the two-time fractional integral reads explicitly
\begin{eqnarray}
\label{corr_eq_fracInt}
\left(\frac{\partial}{\partial t_2}+\frac{\partial}{\partial t_1}\right)^{-\alpha}\left[E_\alpha(-\gamma t_2^\alpha)+E_\alpha(-\gamma t_1^\alpha)\right]&=&\Theta(t_2-t_1)\left[{_0}D_{t_1}^{-\alpha}E_\alpha(-\gamma t_1^\alpha)+\frac{1}{\Gamma(\alpha)}\int_0^{t_1}dt\,t^{\alpha-1}E_\alpha(-\gamma(t_2-t)^\alpha)\right]\nonumber\\
&&+\Theta(t_1-t_2)\left[{_0}D_{t_2}^{-\alpha}E_\alpha(-\gamma t_2^\alpha)+\frac{1}{\Gamma(\alpha)}\int_0^{t_2}dt\,t^{\alpha-1}E_\alpha(-\gamma(t_1-t)^\alpha)\right].
\end{eqnarray}
The second integral can be evaluated with the series representation of the Mittag-Leffler function. We note that
\begin{eqnarray}
\int_0^{t_1}dt\,t^{\alpha-1}(t_2-t)^{\alpha n}=\frac{t_1^\alpha t_2^{\alpha n}}{\alpha}{_2}F_1\left(\alpha,-\alpha n,\alpha+1;\frac{t_1}{t_2}\right),
\end{eqnarray} 
where ${_2}F_1(a,b,c;z)$ denotes the Gaussian hypergeometric function \cite{Abram}.
Consequently
\begin{eqnarray}
\frac{1}{\Gamma(\alpha)}\int_0^{t_1}dt\,t^{\alpha-1}E_\alpha(-\gamma(t_2-t)^\alpha)=\frac{t_1^\alpha}{\Gamma(\alpha+1)}\sum_{n=0}^\infty\frac{(-\gamma t_2^\alpha)^n}{\Gamma(\alpha n+1)}{_2}F_1\left(\alpha,-\alpha n,\alpha+1;\frac{t_1}{t_2}\right).
\end{eqnarray}
Upon substituting this expression as well as Eq.~(\ref{fracInt_ML}) back into Eq.~(\ref{corr_eq_fracInt}), we obtain our second main result from Eq.~(\ref{corr_eq}), the correlation function for initial equilibrium in closed form ($t_2>t_1$)
\begin{eqnarray}
\label{corr_eq2}
\langle x(t_2)x(t_1)\rangle_0&=&\frac{\sigma t_1^\alpha}{\Gamma(\alpha+1)}\sum_{n=0}^\infty\frac{(-\gamma t_2^\alpha)^n}{\Gamma(\alpha n+1)}{_2}F_1\left(\alpha,-\alpha n,\alpha+1;\frac{t_1}{t_2}\right)+\frac{\sigma}{\gamma}E_\alpha(-\gamma t_2^\alpha).
\end{eqnarray}
The limit cases hold as before, as one can easily check. In particular for $\alpha\rightarrow 1$ ($t_2>t_1$)
\begin{eqnarray}
\label{corr_a1}
\lim_{\alpha\rightarrow 1}\langle x(t_2)x(t_1)\rangle_0=\frac{\sigma}{\gamma}e^{-\gamma(t_2-t_1)}.
\end{eqnarray}
With above equations one can see that the CTRW correlation function becomes a function of the time difference $t_2-t_1$ only in the limit $\alpha\rightarrow 1$. This characteristic property is not evident from the autocorrelation function Eq.~(\ref{corr_t0}). Eq.~(\ref{corr_eq2}) therefore clearly demonstrates, that the fractional Ornstein-Uhlenbeck process is a non-stationary process for diverging characteristic waiting times $\langle t(s)\rangle$ of the random walk particle.

Fig.~\ref{Fig_corr} shows the correlation function Eq.~(\ref{corr_eq2}) for different $\alpha$-values. In Fig.~\ref{Fig_corrML} the deviation from a Mittag-Leffler decay is clearly visible. The two curves in Fig.~\ref{Fig_corrML} only agree for $t_1=0$.

\section{\label{sec_con}Conclusion}

We have derived expressions for the correlation function of the fractional Ornstein-Uhlenbeck process, starting from a representation in terms of coupled Lan\-ge\-vin equations. For initial equilibrium position of the random walk particle we were able to obtain the correlation function in analytical form.  This correlation function exhibits two main features. Firstly, it demonstrates the non-stationarity of the CTRW in a harmonic potential for $0<\alpha<1$. Secondly, it clearly deviates from the familiar Mittag-Leffler decay. Both findings are in contrast to the properties of the autocorrelation function Eq.~(\ref{corr_t0}), to which our result converges for $t_1=0$. Considering the widespread application of the single-time quantity Eq.~(\ref{corr_t0}), comparing experimental data with the correlation function Eq.~(\ref{corr_eq2}) should be a worthwhile task. In the end only multi-time statistics are appropriate measures in order to distinguish CTRWs from other non-Markovian processes.

\end{document}